\pdfoutput=1
\documentclass[10pt]{article}
\usepackage{amssymb}

\usepackage{graphicx}
\usepackage{amsmath}

\usepackage{pdfpages}

\input{tcilatex}

\begin{document}

\title{Tolman's Luminosity-Distance, Poincar\'{e}'s Light-Distance and
Cayley-Klein's Hyperbolic Distance\\
}
\author{Dr. Yves Pierseaux (ypiersea@ulb.ac.be)}
\maketitle

\begin{abstract}
(This is a paper of General Physics because it is entirely based on Lorentz
Transformation (LT) but given that we introduce a new definition of
light-distance in Einstein's kinematics, it has cosmological implications)

We deduce Tolman's formula of luminosity-distance in Cosmology from
Poincar\'{e}'s definition of light-distance with LT. In Minkowski's metric,
if distance is proper time (as it is often argued) then light-distance must
be also the shortest... like proper time (unlike Einstein's longest length).

By introducing Poincar\'{e}'s proper light-distance in Einstein's basic
synchronization we deduce a $k-$dilated distance ($k=1+z)$ with relativistic
Doppler's factor) between the observer and the receding ''mirror''.

Such a ''light-luminosity'' distance corresponds not to an Euclidean
distance (Einstein's rigid rod) but to an Hyperbolic Distance (Cayley-Klein)
with a Lobatchevskian Horizon.

From a basic proportionality ''hyperbolic distance- hyperbolic velocity'',
we deduce the law of Hubble. With Hubble's horizon $R_{H}$ we have not only
the constant of Hubble $HR_{H}=c$, but also a minimal Milgrom's acceleration 
$a_{M}R_{H}=c^{2}$. In basic Hyperbolic Rotation (active LT or Einstein's
boost), the former is an hyperbolic angular velocity and the latter a
centrifugal (hyperbolic) proper acceleration.

The cosmological constant is the square $\Lambda =\varrho _{H}^{2}$ of
global Lobatchevskian Curvature $\varrho _{H}=R_{H}^{-1}.$

By following Penrose's Lobachevkian representation of LT, we transform SR
into an Hyperbolic Cosmological Relativity (HCR) by using only the LT but
the whole LT (Einstein's active LT or Einstein's boost).

In Tolman's double reduction, given that photons are dispatching on a sphere
for the source ($R=D$) and the observer ($R=(1+z)D$) as well, the element of
perpendicular area is transformed. We have to take into account LT of solid
angle of aperture of emission light cone and therefore LT of spherical
(isotropic) wavefront into ellipsoidal (anisotropic) wavefront (Poincar\'{e}
1908). The element of perpendicular area (longitudinal ''Lorentz
contraction'') is therefore unchanged. Poincar\'{e}'s elongated
light-ellipsoid becomes a direct explanation of Hubble's expansion.

A non-transversal section in Minkowski's cone is an \textit{ellipsoid} which
transformes pseudo-Euclidean $\pi -relativity$ (steradian is no longer a
invariant), into Lobatchevskian $e-relativity$.
\end{abstract}

\section{Luminosity-distance in Cosmology and invariance of solid angle}

\textbf{(Stationary Source) }The standard method of estimating distances is
given by the relation between observed and estimated luminosity. A
stationary source $S$ having absolute luminosity $L_{S}$ placed at the
distance $D$ will have apparent luminosity $l_{stationary}$ given by the
inverse square law:

\begin{equation}
l_{s}=\frac{L_{S}}{4\pi D^{2}}\qquad \text{\ \ }D:\text{ radius of emission
sphere with center S}  \label{1}
\end{equation}
The absolute luminosity $L_{S}$ is total luminous power (luminous energy per
second) of the source and $l_{stationary}$ is luminous power per unit 
\textit{perpendicular} \textbf{area} intercepted by unit of \textit{solid
angle} (steradian), received at a distance $D$ from the source. If $L_{S}$
can be estimated from a knowledge of the type of the source observed; then
the stationary luminosity-distance $D$ can be calculated since $l_{s}$ is
directly measurable. It is particularly important that the measurement of $%
l_{s}$ involves a small receptor area $dS$ (on the mirror) and therefore a
small solid angle $d\Omega _{stationary}=\frac{dS_{e}}{D^{2}}$ ($e$ for
emitted sphere) .

\textbf{(Receding Source)} At first pointed by Tolman if the source is
receding \textit{in the line of sight} then the luminosity actually observed
will not be $l_{stationary}$ but a reduced value $l_{recession}$ owing to:

* \textbf{The number effect:} the reduction in the number of photons
arriving because of the lengthening of the travel path of a receding source.
this reduces incident radiation by a factor $1+z$ ($z$ is standard spectral
redshift). Tolman and Robertson suppose that the photons received by the
observer $O$ are dispatching on a spherical wavefront (Einstein's spheres).

* \textbf{The energy effect:} the energy of photons arriving is reduced
because redshift lowers their frequency. the reduction of incident radiation
is by a further factor $1+z.$ The reduced value of luminosity $l_{recession}$
on receptor is given by 
\begin{equation}
l_{r}=\frac{1}{(1+z)^{2}}l_{s}  \label{2}
\end{equation}
where $l_{r}=$ $l_{recession}=$ $l_{reception}$ in system \ $K$ of receptor $%
O$ and $l_{s}=l_{stationary}=$ $l_{source}$ in system $K^{\prime }$ of
source $S$. The factor $(1+z)^{2}$ is called the ''Tolman double reduction
factor''\cite{1}. By introducing (1) in (2) we have

\begin{equation}
l_{r}=\frac{L_{S}}{4\pi (1+z)^{2}D^{2}}  \label{3}
\end{equation}
With the invariance of the flux $l_{s}D^{2}=l_{r}D_{r}^{2}$ 
\begin{equation}
l_{r}=\frac{L_{S}}{4\pi D_{r}^{2}}\qquad \text{\ \ }D_{r}=OS_{\text{%
LUMINOSITY}}:\text{ radius of \textbf{r}eception sphere with center S}
\label{4}
\end{equation}
from which Tolman and also Robertson deduce the comoving distance $D$ (given
by simultaneous events: position of the source at the time of observation)
in function of receded luminosity-distance $D_{r}$ between the $S$ (source)
and $O$ (Observer) (\cite{2} \& \cite{3}) \ 
\begin{equation}
OS_{\text{LUMINOSITY}}=D_{r}=(1+z)D\qquad \ \ \ \ \ \ D=\frac{D_{r}}{1+z}
\label{5}
\end{equation}
This is (5) Tolman-Robertson's formula of expanding spherical wavefront in $K
$. There is however a very subtle (\S 3.5) hypothesis in this standard
accepted reasoning. Given that area units are on emitted sphere ($e$) $S_{e}=
$ $D^{2}$ and on received light sphere ($r$) $S_{r}=D_{r}^{2}$, the element
of perpendicular area obviously cannot be an invariant $dS_{e}\neq dS_{r}$
if we have... spheres (center S). So implicit non-relativistic Tolman's
hypothesis is the invariance of element of \textit{solid angle of aperture
of light cone} in both $K^{\prime }$(source) and $K$(observer) ( \cite{4} \& 
\cite{4bis}): 
\begin{equation}
d\Omega _{e}=\frac{dS_{e}}{D^{2}}=d\Omega _{r}=\frac{dS_{r}}{D_{r}^{2}}
\label{6}
\end{equation}
The element of perpendicular area can be an invariant if the sphere of
emission is transformed into... an observed ellipsoid. Let us suppose that
the spherical wavefront of emission (set of simultaneous events) from the
moving source $S$ of $K^{\prime }$ be not transformed into a spherical
wavefront (set of simultaneous events) but into an (elongated) ellipsoidal
wavefront (the events are no longer simultaneous) for the observer in K
(Poincar\'{e} 1908). We will show that, with Lorentz Transformation (LT, 8) 
\begin{equation}
dS_{r}=D_{r}^{2}d\Omega _{r}=dS_{e}=D^{2}d\Omega   \label{7}
\end{equation}
And given that relativistic transformation is $d\Omega $ $=$:$d\Omega
^{\prime }\frac{1}{(1+z)^{2}}$, we will deduce, immediately from (completed)
SR, Tolman-Robertson's law (5). In many papers we showed that Poincar\'{e}'s
ellipsoid $(x,y,z)$ is directly inscribed in $LT$ (Relativity of
Simultaneity). Given that it is an ellipsoid of Revolution, we showed at two
dimensions that Poincar\'{e}'s elongated ellipse $(x,y)$ involves an
original definition of ligth-distance (that is not Einstein's one) (\cite{9,
10, 11}). We prefer here to take as point of departure basic Minkowski's
diagram (and basic Einstein's synchronization) at only one space dimension $x
$. In this case Poincar\'{e}'s ellipse is reduced to only one point but we
will see that one point $M$ (\textit{Fig3-4}) is sufficient for initiating
the Revolution (of ellipse around Ox).

\section{New symmetry duration-distance from Minkowski's calibration
hyperbolas}

Let us consider fundamental hyperbolas along $Ot$ and $Ox$ in Minkowski's
space-time with the axis, $x,$ $t$ of system $K$ (observer) and with light
velocity $c=1$ (\textbf{Fig1)}. The symmetrical scale (or calibration)
hyperbolas determine the space-time units of measure ($x^{2}-t^{2}=\pm 1)$
with the invariance by $LT$ (8) of timelike interval $t^{2}-x^{2}=t^{\prime
2}=T^{2}$ or spacelike interval $x^{2}-t^{2}=x^{\prime 2}=D^{2}$ ( $\gamma
=(1-\beta ^{2})^{-\frac{1}{2}})$(\cite{5}):

\begin{equation}
x^{\prime }=\gamma (x-\beta t)\ \qquad t^{\prime }=\gamma (t-\beta x)\qquad
\ \ \ \ \ \ \ \ \ x=\gamma (x^{\prime }+\beta t^{\prime })\ \qquad t=\gamma
(t^{\prime }+\beta x^{\prime })  \label{8}
\end{equation}
\ The light asymptotes and the standard \textit{hyperbolic rotation }($HR$:
axis $x^{\prime },$ $t^{\prime }$ ''in scissors'') of system $K^{\prime }$
are represented on \textbf{FigA} or \textit{Fig1}\textbf{(}
\cite[http://arxiv.org/abs/0904.3332]{14}). Minkowski's proper duration $T$
(a timelike interval) is determined by the duration between two
events-at-the-same-place $x^{\prime }=0$ at $O^{\prime }$ in $K^{\prime }$,
we define proper distance (a spacelike interval) $D=O^{\prime }P^{\prime }$
by the distance between two ''two events-at-the-same-time'' $t^{\prime }=0$
in $K^{\prime }$ (simultaneous events, see Einstein's synchronization, \S %
3). Observed in $K$, Einstein's dilated duration $t_{\gamma }$ is determined
by ''two events-\textbf{not}-at-the-same-place'' (8), $(0,1)\overset{LT}{%
\rightarrow }(\beta \gamma ,\gamma ),$ and by symmetry, the dilated
''distance'' $x_{\gamma }$ is determined by ''two events-\textbf{not}%
-at-the-same-time'' (8), $P^{\prime }$ $(1,0)\overset{LT}{\rightarrow }%
P(\gamma ,\beta \gamma ):$ 
\begin{equation}
t_{\gamma }=\gamma T\qquad \ \ \text{\ \ \ \ }x_{\gamma }=\gamma D\qquad
\Rightarrow \text{ \ \ }(\gamma ,\gamma )\text{ \ \ \ }\Rightarrow \text{ \ }%
\frac{D}{T}=c=1=\frac{x_{\gamma }}{t_{\gamma }}  \tag{8-1}
\end{equation}
We will show in details with Einstein's synchronization (\S 3) that $T$ is $%
\frac{1}{2}$ light round-trip to travel distance $D$ (in K: $t_{\gamma }$ is 
$\frac{1}{2}$ light round-trip to travel ''distance'' $x_{\gamma }$). We
focus the attention \textbf{FigA} or \textit{Fig3-4}\textbf{(}
\cite[http://arxiv.org/abs/0904.3332]{14}) on light-point $M^{\prime }$ $%
(1,1)\overset{LT}{\rightarrow }M(k,k)$ with $k=\gamma (1+\beta )$ (8-2, $M$
is a point of Poincar\'{e}'s ellipse, note 2, \S 3-5) 
\begin{equation}
\text{\ }t_{k}=kT\qquad \ \ \text{\ \ \ \ }x_{k}=kD\text{ \ \ \ \ }%
\Rightarrow \text{ \ \ \ \ \ \ }\frac{D}{T}=c=1=\frac{x_{k}}{t_{k}} 
\tag{8.2}
\end{equation}
These basic proportions ($\gamma -dilation$ 8-1 \& $k-dilation$ 8-2) are not
possible with Einstein's standard contraction of distance (10).\ We note
that unlike $\gamma -dilation$, $k-dilation$ involves $O^{\prime }M^{\prime }%
\overset{LT}{\rightarrow }\mathbf{O}M$ and therefore a transformation of
proper distance $D$ with respect to $O^{\prime }$ into a ''true'' \textit{%
distance} $x_{k}$ with respect to $O$. Until now our definition 8-2 however
is a purely geometrical definition without physical meaning.

\section{\protect\bigskip Einstein's synchronization and Poincar\'{e}'s
proper light-distance}

Let us now examine in details Einstein's physical procedure of
synchronization \cite{6} in order to define physically the new distance
(8-2).

\subsection{Synchronized clocks, rigid system and rigid rod (Einstein 1905)}

\textbf{(Stationary Mirror).} Let us first consider Einstein's rigid rod $%
x_{M^{\prime }}=O^{\prime }M^{\prime }=L$ at rest in system $K^{\prime }$,
with a light source at $O^{\prime }$ and a mirror at $M^{\prime }$, . A
light signal is emitted from $O^{\prime }$ at $t^{\prime }=0$, it is
reflected in $t^{\prime }=T$ at $M^{\prime }$ and returns to $O^{\prime }$
at $t^{\prime }=2T=2L/c$ $(c=1)$, the ''time out'' $T$ $\ $being equal to
the ''back time'' $T$. Einstein's clock synchronization uses three
successive physical events $1,2,3$: 
\begin{equation}
O^{\prime }(0,0)_{1}\text{\ \ \ \ \ \ \ \ \ \ \ \ \ }M^{\prime }(L,T)_{2}%
\text{ \ \ \ \ \ \ \ }\ O^{\prime }(0,2T)_{3}\qquad \qquad \ \ \ \ \ \ \ \ \
\ O^{\prime }M^{\prime }=M^{\prime }O^{\prime }  \label{9}
\end{equation}
Einstein synchronizes the two clocks at the ends, $O^{\prime }$ \& \ $%
M^{\prime }$, of the rigid rod by defining the simultaneity of two events
''at a distance''. These two simultaneous events $(0,\ T)_{2}\ \ $\&$\ \
(L,T)_{2}$ are however \textit{not explicitly written} in Einstein's 1905
paper. Given that each end of rigid rod $L$ is defined in $K^{\prime }$ for
any time $t^{\prime }$, the length in $K$ is defined by Einstein with
simultaneous ($t=0$) positions of the ends\textit{\ in K} and therefore by
first LT (8)

\begin{equation}
x^{\prime }=\gamma (x-\beta t)\Longrightarrow x_{M^{\prime }}^{\prime
}=\gamma x_{M^{\prime }}(t=0)\Longrightarrow x_{M^{\prime }}=\gamma
^{-1}x_{M^{\prime }}^{\prime }=\gamma ^{-1}L\text{ \ \ \ \ \ \ \ \ }
\label{10}
\end{equation}

This is Einstein's kinematical interpretation of Lorentz contraction $\gamma
^{-1}$.

\subsection{Synchronous distance, abstract system and light-distance
(Poincar\'{e} 1908)}

\textbf{(Stationary Mirror) }Consider now the same situation but without
Einstein's rigid rod (given a priori) and with a single clock in $O^{\prime
} $. The mirror $M^{\prime }$ is at rest in $K^{\prime }$ (a distant
reflecting object) at an unknown distance. A light signal is emitted at $%
O^{\prime }$ at $t^{\prime }=0$, reflected in $M^{\prime }$ in $T$ and
returns to $O^{\prime }$ at $t^{\prime }=2T$ :

\begin{equation}
O^{\prime }M^{\prime }=D\text{(radar)}=\ T\text{(Minkowski) \ \ \ \ \ \ \ \
\ \ \ \ (round trip)}  \label{11}
\end{equation}
The ''synchronous'' distance $D$ may be measured by a single clock in $%
O^{\prime }$ with a $\frac{1}{2}$''round trip'' $2T$ (''two-ways'') signal.
Such a distance (Bondi's radar method, \cite{7}). Until now nothing is
changed because we can replace Einstein's rigid rod $L$ ($\forall t^{\prime
})$ by ''one half light travel time distance''. Suppose now that
''synchronous'' distance $D$ be a proper distance basically defined by the
difference of space coordinates $\Delta x^{\prime }=D$ between two
simultaneous events in $K^{\prime }$($t^{\prime }=T$):

\begin{equation}
\text{\ }(0,\ T)_{2}\text{ \ \ \ and \ \ \ \ \ \ }(D,T)_{2}\qquad \ \ \qquad 
\text{\textbf{FigA}}\ \text{or \textit{Fig3-4}\textbf{: \ \ }}O^{\prime }\
(0,\ 1)_{2}\text{ \ \ \ \ \ \ and}\qquad M^{\prime }(1,1)_{2}  \label{12}
\end{equation}
This definition $t^{\prime }=T$ \ in $K^{\prime }$ is not compatible with
Einstein's definition $t=0$ in K (10) because the simultaneity is relative:
if both ends of such a distance (12) are given at the same time $T$ in $%
K^{\prime }$, they cannot be determined at the same time in $K$ (10). We add
here a new element because the reciprocal (Poincar\'{e}) interpretation of
Einstein's synchronization involves that exactly as ''simultaneity at a
distance'' cannot be defined without the velocity of light, the \textit{%
distance itself} cannot be defined without this velocity. This \textit{proper%
} \textbf{light-distance} $D$ then becomes an ''\textit{invariant''} of LT ($%
x^{2}-t^{2}=D^{2}=s^{2})$ in the same sense as the proper duration $T$ is an
''invariant'' of LT ($t^{2}-x^{2}=T^{2}=s^{2})$. \textit{In summary Poincar%
\'{e}'s proper distance (hyperbola along }$Ox$\textit{) is the exact
symmetric of Minkowski's proper time (hyperbola along (Ot).} Poincar\'{e}'s
interpretation of Lorentz contraction involves that\textit{\ }proper
distance $D$, like proper time, is the \textbf{shortest} in $K^{\prime }$
(in Einstein's (10), it is the longest in $K^{\prime }$). \textit{''This
Lorentz hypothesis is the immediate translation of Michelson's experiment,
if the lengths are defined by the time that light takes to travel through
them''}(\textit{\cite{8}).}As a last analysis and from a historical
viewpoint, Einstein's work is based on the direct theorem (the $O^{\prime
}t^{\prime }$ axis), while Poincar\'{e}'s opened the way to the reciprocal
(the $O^{\prime }x^{\prime }$ axis).

\subsection{\protect\bigskip Poincar\'{e}'s $k-dilated$ round-trip
light-distance ''Observer-(receding)Mirror''}

\textbf{(Receding Mirror).}Let us examine now the light-distance from system 
$K$ where the mirror is receding from O. Travel-duration in K is given by
the difference of time coordinates $\Delta t=2\gamma T$ between two
not-at-the-same-place events,\ \ \ 
\begin{equation}
O^{\prime }(0,0)_{1}\ \ \&\ \ \ O^{\prime }(0,2T)_{3}\overset{LT}{%
\rightarrow }(0,0)_{1}\ \ \&\ \ (2\gamma \beta T,2\gamma \beta T)_{3}\qquad
\Delta t=2\gamma T=2t_{\gamma }\qquad \text{travel-duration}  \label{13}
\end{equation}
Automatically (11) involves that light-''distance'' $O^{\prime }M^{\prime }$
in $K$ is given by $\frac{1}{2}\Delta t=x_{\gamma }$ (8-1)$.$ Obviously the
one-way \ $O^{\prime }M^{\prime }$ in $K$ is also given by the \textit{%
difference of x coordinates }$\Delta x=\gamma T$ between two not
at-the-same-time events 
\begin{equation}
O^{\prime }(0,0)_{2}\text{ \ \ }\&\text{\ \ }M^{\prime }\text{\ }(D,0)_{2}%
\overset{LT}{\rightarrow }(0,0)_{2}\text{ \ }\&\text{\ \ \ }(\gamma D,\gamma
\beta D)_{2}\qquad \Delta x=\gamma D=x_{\gamma }\qquad \text{%
light-''distance''}  \label{14}
\end{equation}
\ If distance is really time (see Penrose, conclusion), the so formed
distance $D$ must be LTed (8) into $\gamma D$ like the duration.
Unfortunately such a $\gamma -dilated$ ''distance'' (14) is not a physical
distance because\ until now we have only considered $O^{\prime }M^{\prime }$
but not the distance between the receding mirror and \textit{the observer }$O
$: We have only deduced round-trip\ light ''distance'' $O^{\prime }M^{\prime
}O^{\prime }$ in $K^{\prime }$ (11) and in $K$ (14) but not round-trip light
distance $OMO$ in $K$. With our new symmetry duration-distance we can
calculate $OMO.$\ We know that the signal is in $O^{\prime }$ in $2\gamma T$
(13)$\ $and that at this $K-time$ the $K-distance$ $OO^{\prime }$ is $2\beta
\gamma T$ (13).\ We deduce the total travel duration $2\gamma T$ $+2\beta
\gamma T$ . Automatically (11) involves that light-distance $OM$ in $K$ 
\begin{equation}
OM=\sqrt{\frac{1+\beta }{1-\beta }}D  \label{15}
\end{equation}
In the case of stationary mirror (11) and receding mirror (15) the
light-distance is given by half total travel duration as well. This $%
k-dilated$ light distance is Tolman's luminosity distance except that
Tolman's distance is a one-way distance OS (see \textbf{FigA }or \textit{%
Fig3-4,} point $M(x_{k},t_{k}))$.

\subsection{Poincar\'{e}'s one-way $k-dilated$ light-distance, Tolman's
luminosity-distance and relativistic ''Doppler'' formula}

\bigskip Let us consider in details the LT of Einstein's first two (1 and 2)
successive events $O^{\prime }M^{\prime }$ 
\begin{equation}
O^{\prime }(0,0)_{1}\ \ \ \&\ \ M^{\prime }(D,T)_{2}\text{ \ }\overset{LT}{%
\rightarrow }(0,0)_{1}\ \ \ \ \ \&\text{ \ \ \ }(\gamma (D+\beta T),\text{ }%
\gamma (T+\beta D))_{2}\text{\ \ \ \ \ \ }  \label{16}
\end{equation}
The one-way forth light-distance $\Delta x$ is given by difference (final
and initial) of space coordinate $\Delta x=x_{f}-x_{i}$ or time coordinate $%
\Delta t=t_{f}-t_{i}$ as well (light-distance=travel-duration) 
\begin{equation}
\Delta x_{forth}(O^{\prime }\text{ }\rightarrow \text{ }M^{\prime })=\gamma
(1+\beta )T=\sqrt{\frac{1+\beta }{1-\beta }}D  \label{17}
\end{equation}
This is the point (\textbf{FigA }or\textbf{\ }\textit{Fig3-4}) $M^{\prime }$ 
$(1,1)\overset{LT}{\rightarrow }M(k,k)$. With (2 and 3) successive events $%
M^{\prime }O^{\prime }$ the one-way back light-''distance'' $\Delta x$ is
given by difference (final and initial) 
\begin{equation}
M^{\prime }(D,T)_{2}\ \ \&\ \ O^{\prime }(0,2T)_{3}\text{ \ }\overset{LT}{%
\rightarrow }(\gamma (D+\beta T),\text{ }\gamma (T+\beta D))_{2}\text{\ \ }\&%
\text{\ \ \ }(2\gamma \beta T,2\gamma T)_{3}\   \label{18}
\end{equation}
of space coordinate $\Delta x=x_{f}-x_{i}$ or time coordinate $\Delta
t=t_{f}-t_{i}$ as well

\begin{equation}
\Delta x_{back}(M^{\prime }\rightarrow \text{ }O^{\prime })=\gamma (1-\beta
)T=\sqrt{\frac{1-\beta }{1+\beta }}(-D)  \label{19}
\end{equation}
We see that if $\Delta t$ is always positive, that is not the case for
algebraic $\Delta x:$ we have $T=D$ if the travel light is in positive $Ox$
sense and $T=-D$ in negative sense (note 3). We rediscover\ $(\sqrt{\frac{%
1+\beta }{1-\beta }})_{O^{\prime }M^{\prime }}+(\sqrt{\frac{1-\beta }{%
1+\beta }})_{M^{\prime }O^{\prime }}=2\gamma $ and given that $\Delta
x_{oneway}(O^{\prime }$ $\rightarrow $ $M^{\prime })\neq \Delta
x_{oneway}(M^{\prime }\rightarrow $ $O^{\prime })$, the round-trip
definition of distance seems to be in physics the only one possible\footnote{%
But imagine a physical situation where the round trip is by definition
impossible (in\ Cosmology).}. However until now we have only considered
one-way $O^{\prime }M^{\prime }$ and $M^{\prime }O^{\prime }$ in $K^{\prime
} $ and in $K$ but not the observer $O$. According to oberver $O$, the
receding mirror is $M$. Let us consider now one-way light \textit{distance} $%
OM$ and $MO$. In the first case (16) $OM$ is given by (17) the light travel
duration from the emission in $O$ and the reception in $M$ and therefore $%
OM=\Delta x_{oneway}(O^{\prime }$ $\rightarrow $ $M^{\prime })$. In the
second case $MO$ is not given by (19, $MO\neq \Delta x_{oneway}(M^{\prime
}\rightarrow $ $O^{\prime })$) but by light travel duration from the
reflection in $M$ and the reception in $O$ (and not $O^{\prime },$19). Given
that light signal is in $O^{\prime }$ at $\gamma (1+\beta )T+\gamma (1-\beta
)T$ $=2$ $\gamma T$ (18)$\ $and that at this $K-time$ the $K-distance$ $%
OO^{\prime }$ is $2\beta \gamma T.$(18) we have a consistent new definition
of $k-$\textit{dilation} light-distance with $\gamma (1-\beta )T+2\gamma
\beta T$, 15). 
\begin{equation}
OM=MO=\sqrt{\frac{1+\beta }{1-\beta }}D=kT  \label{20}
\end{equation}
In summary, given that source and mirror are at rest in $K^{\prime }$ and
that $O^{\prime }$ and $O$ coincide in $t=t^{\prime }=0$, we have Einstein's
equality of one way travel time $O^{\prime }M^{\prime }=M^{\prime }O^{\prime
}$ and Poincar\'{e}'s equality $OM=MO$ as well\footnote{$M$ is the only
(right) point of Poincar\'{e}'s elongated ellipse in K with the observer $O$
at the (left) focus and the source at the center $O^{\prime }.$ (\cite{9})}
\ This is a physical new definition of distance if we reverse the situation
''source-mirror'' in K': the remote source $S$ is now at proper distance $D$
in $K^{\prime }$ at $t=t^{\prime }=0$ when $O$ and $O^{\prime }$ coincide
and the mirror of the telescope is in $O$. What is the cosmological
light-distance $OS$ or $MS?$ . Two events, ''coincidence'' $O\equiv
O^{\prime }$ $(0,0)$ and ''emission'', $(D,0)$ are simultaneous\footnote{%
According to a non-relativistic calculation (absolute simultaneity) the
light signal catch up with $O$ at the distance$\frac{D}{1-\beta }.$ In
relativistic point of view we have rigorously proved (15 and 20, the
principle of inverse return of the light) that we have $\sqrt{\frac{1+\beta 
}{1-\beta }}D$.} in $K^{\prime }$ but not in K $(0,0)$ and $(\gamma D,\gamma
\beta D)$. Then the time of emission $t_{i}=t_{e}$ is not the same in $%
K^{\prime }$ and $K.$ Given that the signal is in $O^{\prime }$ in $\gamma T$
and the distance $OO^{\prime }=\gamma \beta T,$ the total duration until $O$
gives$\ \gamma T+\gamma \beta T$ $=\gamma \tau T(1+\beta )=\sqrt{\frac{%
1+\beta }{1-\beta }}T\ $. We obtain now the identity between \textbf{%
Poincar\'{e}'s light-distance} and \textbf{Tolman's luminosity-distance }$%
D_{r}$\textbf{\ }(5).

\begin{equation}
OS_{\text{LIGHT}}=\sqrt{\frac{1+\beta }{1-\beta }}D=D(1+z)=OS_{\text{%
LUMINOSITY}}  \label{21}
\end{equation}
where $D$ is a proper or a comoving distance. Unlike $\gamma -dilated$
distance (14), $k-dilated$ distance is a physical distance.We can also
obtain this basic result with $t_{r}-t_{e}>0$ by taking into account the
negative sense of travel light. $D=-T$ : $S(-T,0)_{i}\ \ \ \&\ \ O^{\prime
}(0,T)_{f}$ \ $\overset{LT}{\rightarrow }(-\gamma T,-\gamma \beta T)_{e}\ \&$
\ $(\gamma \beta D,\gamma T)_{f}$ and therefore (21, with the same method we
transform 19 in 20).\ 

Suppose now that the proper light-distance $D$ of a very remote
monochromatic source $S$\ $\ $be unknown and only the length-wave $\lambda
_{S}$ and the intensity of source $L_{S}$ (absolute luminosity) are known.\
We have then a basic proportionality distance-lengthwave with a new redshift 
\textbf{Light-Luminosity Distance}

\begin{equation}
\frac{D_{r}}{D}=\sqrt{\frac{1+\beta }{1-\beta }}=\text{\ }\frac{\lambda _{%
{\small Obs}}}{\lambda _{Source}}=k\ =1+z  \label{22}
\end{equation}
This is obvious because Poincar\'{e}'s length is basically a travel length
by a wave. So we deduce Tolman-Robertson's law (5), coupled with
relativistic ''Doppler\footnote{%
CR is based not on plane wavefront but on spherical wavefront (\cite{11})}''
redshift law (22), directly from LT and new definition of distance
(inscribed in Minkowski's diagram, \textbf{FigA} or \textit{Fig1-3}). This
is the reason why we suggest to call the new SR with Cosmological Relativity
(\textbf{CR}). The $k-$dilation (21 or 22) is a law of expanding universe.
We note however that Tolman-Robertson's law (5) is based on spherical waves
and thus on an area on the mirror of the telescope: we have now to prove
that this law is immediately deductible from LT at 3 dimensions.

\subsection{Tolman's double reduction, Double Reduction of solid angle of
emission and Poincar\'{e}'s space-time light ellipsoid}

Light-distance by travel-duration can be generalized at $3$ space
dimensions; $\vec{r}$($x,y,z)$ \& $\vec{r}^{\prime }$($x,y,z),$ with the
norms $r=t$ and $r^{\prime }=t^{\prime }(c=1)$ and with a source emitting a
spherical wavefront in $K^{\prime }$. Given that the azimutal angle is
Lorentz invariant (ellipsoid of revolution, see \S 1), we consider only the
angle $\theta $ and $\theta ^{\prime }$(respectively in $x,y$ and $x^{\prime
},y^{\prime }$ planes) and solid angle of the light cone of emission $\Omega
=2\pi (1-\cos \theta )$ LTed into $\Omega ^{\prime }=2\pi (1-\cos \theta
^{\prime })$.\ 
\begin{eqnarray}
x &=&\gamma (x^{\prime }+\beta t^{\prime })\ \ \ \ \ \ \ \ \ \ \ \ \ \ \ \ \
\ \ \ \ \ \ y=y^{\prime }\qquad \text{\ \ \ \ \ \ \ \ \ \ \ \ \ }t=\gamma
(t^{\prime }+\beta x^{\prime })\text{ }  \label{23} \\
r\cos \theta  &=&\gamma (r^{\prime }\cos \theta ^{\prime }+\beta r^{\prime
})\ \text{\ \ \ \ \ \ }r\sin \theta =\text{\ }r^{\prime }\sin \theta
^{\prime }\text{\ \ \ \ \ \ \ \ \ \ }r\text{\ }=\gamma r^{\prime }(1+\beta
\cos \theta ^{\prime })  \label{24}
\end{eqnarray}
and we rediscover Einstein's aberration formula $\cos \theta =\frac{\cos
\theta ^{\prime }+\beta }{1+\beta \cos \theta ^{\prime }},$ Penrose's
formula $\tan \frac{1}{2}\theta =k^{-1}\tan \frac{1}{2}\theta ^{\prime }$
and polar equation of Poincar\'{e}'s ellipse $r=t=$ $\frac{\rho }{\gamma
(1-\beta \cos \theta )}$ where $\rho $ corresponds to $D$ ($M$ $\in $
ellipse, \textbf{FigA} or \textit{Fig3-4}, note 2, \cite{9}). The isotropic
(spherical) emission of a moving source ($S$ in $K^{\prime }$) is
anisotropic (ellipsoidal\footnote{%
Poincar\'{e}'s ellipsoid of revolution is the direct kinematical explanation
of the very physical \textit{''headlight effect'' } (in synchrotron
radiation, bremsstrahlung...) The LTed\textit{\ ellipsoidal wavefront is an
equiphase surface} (a non-transverse section in Minkowski's cone)}) observed
from $K$ with relativistic transformation of the solid angle (\cite{8})

\begin{equation}
\Omega =\Omega ^{\prime }\frac{1}{\gamma ^{2}(1+\beta \cos \theta ^{\prime
})^{2}}\   \label{25}
\end{equation}
This is reduction of the angle of aperture of the cone of emission of a
moving source. For small angle $\theta ^{\prime }$ we have no aberration
(Robertson, the motion is along \textit{the line of sight}) but a headlight
effect (Lorentz reduction of solid angle given that we have necessarily a
small area on the mirror of telescope):

\begin{equation}
d\Omega =\ d\Omega ^{\prime }\frac{1-\beta }{1+\beta }=d\Omega ^{\prime
}k^{-2}\text{ }=d\Omega ^{\prime }\frac{1}{(1+z)^{2}}\text{ \ \ \ \ \ \ \ \ }%
r\text{\ }=r^{\prime }\sqrt{\frac{1+\beta }{1-\beta }}=k\rho \text{\ }%
=(1+z)\rho \text{\ \ \ \ \ \ \ \ }(d\theta =k^{-1}d\theta ^{\prime })
\label{26}
\end{equation}
Where $r$ corresponds to $D_{r}$ (20 or 21). And so we\ deduce directly the
law of Tolman-Robertson from new fundamental relativistic invariant in 
\textbf{CR} the element of ''perpendicular'' area must be an invariant
(purely longitudinal\footnote{%
According to Poincar\'{e}, the ellipsoid is elongated because units, meters
and \textit{steradians} are reduced \cite{8}. Of course the right angle of
perpendicular area is LTed.} elongation of Poincar\'{e}'s ellipse, \cite{10}%
) 
\begin{equation}
dS=r^{2}d\Omega =dS^{\prime }=r^{\prime 2}d\Omega ^{\prime }\qquad \qquad 
\frac{d\Omega }{d\Omega ^{\prime }}=\frac{l_{recession}}{l_{stationary}}
\label{27}
\end{equation}
Poincar\'{e}'s \textit{double} reduction of angle of aperture of emission
cone (26) is exactly Tolman's \textit{double} reduction of luminosity (2).
Poincar\'{e}'s space-time elongated light ellipsoid\footnote{%
Let us remark that Robertson, in order to prove Tolman's formula, uses
Einstein's 1905 formula of LTed volumic density of energy in Complex of
Light. But Einstein's spherical Complex is also LTed into an ellipsoid ( 
\cite[§8]{6}).} is therefore a direct explanation of Hubble's expansion
(Observer is at the focus of the meridian section of ellipsoid and therefore
the geometrical measure of solid angle, steradian, is not a Lorentz
invariant). We must prove now that Poincar\'{e}'s light-distance necessarily
involves an Lobatchevskian distance with an Horizon.

\section{Hyperbolic velocity, hyperbolic distance and hyperbolic Hubble's law%
}

In Friedman-Lema\^{i}tre's model (in Robertson-Walker's metric), the
theoretical Hubble law is defined by an apparent velocity $V_{\exp }$ of
expansion of geometrical space itself $R(t),$ the scale factor, that is not
limited by the velocity of light$\ \ $%
\begin{equation}
V_{\exp }=\dot{R}(t)\ =H(t)R(t)  \label{28}
\end{equation}
In standard model $RW$ $\ $the ''constant'' $H(t)$ of Hubble is defined by
its present value $H$ and the law of expansion is not connected with
relativistic Doppler's formula in $SR$ (scale factor $k$). However, the
experimental measurements are made not on the space itself but on the moving
bodies. So the empirical form of Hubble's law is a relation between spectral
redshift $z$ and distance $\rho $ (or $D$) of remote objects; deduced from
Tolman's law (5) generally written 
\begin{equation}
v=cz=H\rho \ \ \ \ \ \ \ \ \ \ \ \ \frac{\lambda _{observer}}{\lambda
_{source}}=1+\frac{v}{c}=1+z  \tag{28bis}
\end{equation}
with non-relativistic Doppler law (\cite{2}). The ''constant''\footnote{%
The cosmologists introduced a magnitude without dimension $h_{0}$ a fraction
of $100km/sec/Megaparsec$ ($h_{0}=0.5?$).} of Hubble, that is defined by
this empirical law is directly confirmed when the redshift shift is small
compared with unity $v<<c$. When this is not the case, for example quasar
3C9 a wave length ratio of $3.01$ ($z=2.01$) for which $v>c$ a correction
with Einstein's Doppler law is necessary $1+z=k$ $\Rightarrow v/c=0.8.$
Thanks to this correction on velocity $v$, we have $\rho <present$ $R(t)$
with $present$ $R(t)=\frac{c}{presentH(t)}.$ So we have the paradox in
standard $RW$ that (28) has nothing to do with SR (non-Minkowskian metric)
whilst its experimental form (28bis) is directly connected with SR but only
with spectral lenghtwave $\lambda $ (not for length ''itself'' $\rho $). And
we showed that in CR lengthwave and length are LTed in the same way (22).

How can we deduce rigorously a basic law of Hubble, i.e. a basic
proportionality between $\beta $ and $\rho $ in CR (with LT)? \ The equation
\ $r$\ $=\rho \sqrt{\frac{1+\beta }{1-\beta }}$ (26) or (21) suggests an 
\textit{hyperbolic definition of Poincar\'{e}'s light distance} in the
meaning of Cayley and Klein. In Beltrami's model of hyperbolic geometry : a
circle of radius $R_{H}$ is regarded as an horizon (a circle ''at
infinity'') and a straight line is interpreted as a line segment within this
circle. Cayley and Klein define an hyperbolic distance by the \textit{%
cross-ratio formula}. Consider the hyperbolic radial distance $r_{H}$ from
origin of the circle to a point $P$ with Cartesian distance $\rho :$ 
\begin{equation}
r_{H}=\frac{R_{H}}{2}\ln \text{ }\frac{R(R+\rho )}{R(R+\rho )}\text{\ \ \ \ }%
\Rightarrow \text{\ \ \ }r_{H}=R_{H}\func{arctanh}\frac{\rho }{R_{H}}%
=R_{H}\ln \sqrt{\frac{1+\frac{\rho }{R_{H}}}{1-\frac{\rho }{R_{H}}}}
\label{29}
\end{equation}
By taking the Neperian logarithm (number $e$) of the fundamental formula
(22) it turns out $(c=1)$: 
\begin{equation}
\beta _{H}=\ln \sqrt{\frac{1+\beta }{1-\beta }}=\ln \frac{\lambda _{{\small %
Obs}}}{\lambda _{Source}}=\ln k\ =\ln (1+z)=Z  \label{30}
\end{equation}
where $\beta _{H}$ is the hyperbolic velocity in $SR$. So if we compare (29
and 30), we have \textbf{a fundamental Hyperbolic proportionality between }$%
\beta _{H}$\textbf{\ and }$r_{H}$ if and only if\ $\frac{\rho }{R_{H}}=\beta 
$ \ ($\Longrightarrow k\beta =Hr).$We note $Z$ \ as the logarithmic spectral
redshift\footnote{%
Given that $Z$ is a strictly increasing function of the wavelength ratio
which is zero when $\frac{\lambda _{observer}}{\lambda _{source}}=1$, $%
Z\approx z$ when $z$ is small and for infinitesimal wavelength shift we have 
$\delta Z=\delta z=\frac{\delta \lambda }{\lambda }$ (logarithmic derivation)%
$.$} (\cite{4}):.

\begin{equation}
\beta =H\rho \qquad \qquad \beta _{H}=Hr_{H}=Z\text{ \ \ \ \ \ \ \ \ \ \ \ \
\ \ \ \ \ \ \ \ \ \ }\ (c\beta =Hr\qquad c\beta _{H}=Hr_{H})  \label{31}
\end{equation}
hyperbolic or non-hyperbolic expression of Hubble's law which becomes a
basic law of CR completely defined by LT (8) and therefore by Hyperbolic
Rotation $HR$ ($\beta _{H}$ is hyperbolic angle of rotation). Given that in
any Rotation motion, there is an acceleration, Hubble's constant appears as
a basic Hyperbolic Angular Velocity (Euclidean angular velocity $v=\omega R)$%
. Let us examine if Hubble constant correspond to a basic Hyperbolic
Acceleration ($Hc=\alpha _{M}=\frac{c^{2}}{R_{H}})$ (\cite{14}). Suppose
Einstein's basic boost where $K^{\prime }$ ($d\tau $ element of proper time)
is uniformly accelerated (from $0$ to $\beta _{H})$ with respect to $K$

\begin{equation}
Hc=\alpha _{M}=\frac{c^{2}}{R_{H}}\qquad \overset{c=1}{\Longrightarrow }%
\qquad \frac{d\beta _{H}}{dt}=\gamma ^{2}\frac{d\beta }{dt}\text{\qquad }%
\frac{d\beta _{H}}{d\tau }=\gamma ^{3}a=\dot{\beta}_{H}=\alpha _{M}\qquad
(c=1)  \label{32}
\end{equation}
$\alpha _{M}$ being a minimal \textit{non-null }norm of spacelike 4-vector
of acceleration ($a$ can be as small $\gamma ^{-3}$ as we wish). In standard 
$SR$, given that $HR$ is not a motion, we have for active $LT$ or $HR:$ $%
\frac{d\beta _{H}}{d\tau }=0$.\ Suppose now that $HR$ (8) be a physical
motion (Born-Rindler's hyperbolic ''rigid'' motion,  (\textit{Fig2}, \cite
{14}) we have 
\begin{equation}
\frac{d\beta _{H}}{d\tau }=H=\frac{\dot{k}}{k}=\frac{1}{1+z}\frac{\delta z}{%
\delta \tau }\   \label{33}
\end{equation}
In standard static metric we have obviously $k=const$. Recent observations
indicate a variation $\delta z$ in our proper time $\delta \tau $. In
Cosmology we measures always $z$ and $D_{r}$ but never $R(t)$. This ad hoc
scale factor in RW's metric with an absolute time $t$ can be eliminated with
Occam's razor in HCR (Hyperbolic Cosmological Relativity).. By integration
with standard initial conditions of basic Lorentz boost ($O\equiv O^{\prime
},t=t^{\prime }=0)$

\begin{equation}
d\beta _{H}=Hd\tau \text{ \ \ \ \ \ }\Longrightarrow \text{\ \ \ }\beta
_{H}=H\tau \text{\ \ }\Longrightarrow \text{\ \ \ }r_{H}=\tau \text{ \ }(c=1)
\label{34}
\end{equation}
We deduce the physical relativistic meaning of $r_{H}:$ hyperbolic distance
is proper time $\tau c$ (see Penrose, conclusion) in a basic Einstein's
boost or basic HR (where $H$ is the angular velocity). We have used only the
LT and the whole LT (HR: passive-LT but also ''active-LT or Einstein
boost'').

\section{Conclusion: From Penrose's Lobatchevskian SR to Cosmological
Hyperbolic Relativity}

The question now is: how can we modify or adapt standard SR to the new
symmetrical hyperbolic light-distance? Nothing is changed about proper time
(and therefore about standard relativistic dynamics). Penrose writes about
SR and Hyperbolic geometry:

\begin{quotation}
I said that I like hyperbolic, Lobatchevskian geometry the best. One of the
reason is that the group of symmetries is exactly the same as.. the Lorentz
group, the group of SR.\ \ \ (...) Distance in Minkowskian geometry is time,
the proper time that is physically measured by moving clocks. It turns out
that the intrinsic geometry of the ''sphere'' (in Minkowskian space-time) is
Lobatchevskian hyperbolic geometry \cite{15}.
\end{quotation}

\bigskip It could be argued that with a rigorous definition of
light-distance by proper time, nothing is changed with standard Einstein's
asymmetrical contraction of distance (''Gedanken experiment'' never
experimentally observed). But if distance is proper time (\S 3), it must be
dilated, in rest frame like... proper time. We showed that Einstein's
(contracted) rigid rod is no longer valid for Cosmological distances (in
light-years). If distance is proper time (from O' in K'), light-distance
(from O in K) must be $k-dilated$ with travel duration (21). Such a distance
determines not an Euclidean rigid rod by an Hyperbolic distance. We showed
finally that Hyperbolic Distance $r_{H}$ is directly proportional to
Hyperbolic Velocity $\beta _{H}$ (law of Hubble, 31) but also that $r_{H}$
is proper time $\tau $ (34)\ in elastic motion (Born-Rindler ''rigid''
motion without Einstein's rigid rod).

Unlike Galilean invariant with Euclidean distance defined by plus $(+)$
signs $r^{2}=x^{2}+y^{2}$, Lorentz invariant involves one minus $(-)$ sign.
So a standard objection could be that SR is already hyperbolic because (at
one dimension for example, \textbf{FigA}) we have a minus sign in particular
for scale hyperbola along Ox $x^{2}-t^{2}=x^{\prime 2}(t=0)$. Standard SR
(signature: $(1,-1)$ is not completely Lobatchevskian because Lobatchevskian
geometry involves necessarily an Horizon $x^{\prime }=R_{H}$ (and a
curvature $\varrho _{H}=\frac{1}{R_{c}}$), unlike Euclidean geometry which
involves $x^{\prime }\rightarrow \infty $ (without horizon and flat $\varrho
_{E}=0$). So when all physicists, during more than one century, write

\begin{equation}
\text{(LEFT Member: Hyperbolic) \ \ \ \ \ \ \ }x^{2}-t^{2}=x^{\prime
2}\nrightarrow \infty \text{ \ \ \ \ \ \ \ \ (RIGHT Member: Euclidean, Flat)}
\label{35}
\end{equation}
they introduce an Euclidean definition of infinity in the second member of
Minkowski's basic invariant: this Euclidean flatness is a ''stranger'' in an
hyperbolic interval $(1,-1)$. So they obtain the standard \textit{flat}
pseudo-euclidean geometry. What does it mean physically? Minkowski claimed,
that \textit{''space by itself and time by itself are doomed to fade away
into mere shadows, and only a kind of union of the two will preserve an
independent reality''}. But an ''infinite'' interval\footnote{%
If we delete scale hyperbolas we have a non-relativistic infinite interval
and a flat space-time} ($x^{2}-t^{2}$ \ \ $=$ \ $\infty ,$ $\forall t$)
should mean that independent space is given for any $t$ and therefore the
Return of the Shadow (Absolute Space $Ox$, $\forall t$). So the
pseudo-Euclidean flatness promotes the Return of the Phantom (Remake II: the
dark energy\footnote{%
It is a remake of Absolute physics because dark matter is not dark energy! ($%
E=mc^{2}$, Einstein 1905)}?). It is therefore \textbf{logically }essential
to put an horizon in (35).

\begin{equation}
\text{Hyperbolic }(1,-1)\text{ \ \ \ \ \ \ \ \ }x^{2}-t^{2}=R_{H}^{2}=%
\varrho _{H}^{-2}\ \ \ \ \ \ \ \ \text{Horizon }R_{H}\text{, Global negative
Curvature }\varrho _{H}  \label{36}
\end{equation}
Only (36) is consistent. This recalibration is also a renormalization of
Minkowski's metric because we eliminate Euclidean inappropriate infinity. In
this way, we transform pseudo-Euclidean SR into Hyperbolic Cosmological
Relativity (\textbf{HCR}).

We rediscover our equation (2) of our $\Lambda $-paper (\cite[equation 2]{14}%
) because Cosmological Constant $\Lambda =\varrho _{h}^{2}$ is the square of
Lobatchevskian basic global (negative) Curvature .Given that LT is an
Hyperbolic Rotation (\textbf{FigA}) and that in any Rotation, there is an
acceleration (Einstein's boost) we showed, from Born-Rindler's accelerated
''rigid'' (?) motion (\cite{12b}), that Hyperbolic Rotation Motion is an 
\textit{elastic}\footnote{%
We will show in CEMB that the elasticity constant is $G^{-1}($the inverse of
gravitational constant in units $c=1:g/cm$).} motion involving a \textit{%
centrifugal\footnote{%
Euclidean Rotation Motion involves a \textit{centripetal} acceleration}}
acceleration (coupled with expanding distance). With the horizon of Hubble,
we deduce a basic Hyperbolic Acceleration in basic Einstein's boost that is
Milgrom's minimal \textit{proper} acceleration $\alpha _{M}=\frac{x^{2}}{%
R_{H}}$ (MOND\footnote{%
In HCR we have therefore a basic proportionality with the inverse of
distance and not with the inverse-square of distance.},\cite{12}). Hubble
constant is a basic Hyperbolic \textit{Angular }Velocity ($Hc=\alpha _{M}).$
We rediscover our equation (4) in $\Lambda $-paper

\begin{equation}
x^{2}-c^{2}t^{2}=\frac{c^{4}}{\alpha _{M}^{2}}=R_{H}^{2}=\mathbf{\Lambda }%
^{-1}.\text{\ \ \ \ \ \ }\Longrightarrow \text{\ \ \ \ \ \ \ }\frac{1}{%
R_{H}^{2}}=\varrho _{H}^{2}=\text{ \ }\mathbf{\Lambda }\text{ \ }=\text{\ }%
\frac{\alpha _{M}^{2}}{c^{4}}=\frac{H^{2}\text{ }}{c^{2}}\text{\ \ }(c=1) 
\tag{4-$\Lambda $}
\end{equation}
In HCR true constants $R_{H},$ $H,$ $T_{H}$ are directly connected to
constant $\mathbf{\Lambda .}(c=3.10^{10}cm/s$, $H\approx 10^{-18}s,$ $%
R_{H}\approx 3.10^{28}cm,\alpha _{M}\approx 3.10^{-8}cm/s^{2}$ $\Lambda
=10^{-57}cm^{-2}).$ Can we expect with hyperbolic distance a small
adaptation or a very large adaptation of SR? In one sense it is a small
adaptation because Minkowski's signature is unchanged. The
''renormalization'' (recalibration) of standard flat element of Minkowski's
metric is

\begin{equation}
dt^{2}-dx^{2}\text{ }=d\tau ^{2}=\alpha _{M}^{-2}d\beta _{H}^{2}=\varrho
_{H}^{-2}d\beta _{H}^{2}\text{ }  \tag{34-$\Lambda $}
\end{equation}
\ But in another sense it is a very big adaptation because we does not need
in Cosmology the standard factor of scale $R(t)$ (28) in $RW$'s metric (we
underline that hyperbolic metric for $\kappa <0$ in $RW$ obviously is not
Minkowskian). So the standard hypothesis (Lema\^{i}tre-Gamow-Paccelli) $%
R(t)=0$ can be replaced by a well-tempered (in proper time $\tau $) ''steady
state'' infinite (Hyperbolic) space-time (Hoyle-Bondi).

On an historical point of view there are two curious ideas in Poincar\'{e}'s
work with respect to Einstein's standard SR: the light ellipsoid
(kinematics) and the negative \textit{gravitational} pressure of classical
vacuum (dynamics). Poincar\'{e} unfortunately has had no enough time to
develop both ideas that are today completely forgotten. HCR is a kinematical
synthesis between Einstein's kinematics (5) and Poincar\'{e}'s light
distance (8). In our next paper we will develop the corresponding dynamics
and more precisely ''ELECTRO-dynamics''. We will prove that in CEMB
(Cosmological ElectroDynamics of Moving Bodies), Poincar\'{e}'s 1908
(negative) pressure of classical vacuum (\cite{13}) and Einstein's 1917
cosmological constant are directly connected.

The irony of history is that the complete \textit{hyperbol}ization of SR
(30) is induced by a non-transversal section in Minkowski's cone an
therefore by an... \textit{ellipsoid} which transformes pseudo-euclidean $%
\pi -relativity$, given that (ste)radian is no longer an invariant, into
Lobatchevskian $e-relativity.$(Penrose). According to Poincar\'{e}, in
french dans le texte ''les cercles divis\'{e}s dont nous nous servons pour
mesurer les angles sont d\'{e}form\'{e}s par la translation, ils deviennent
des ellipses''(\cite{8}).

\includepdf{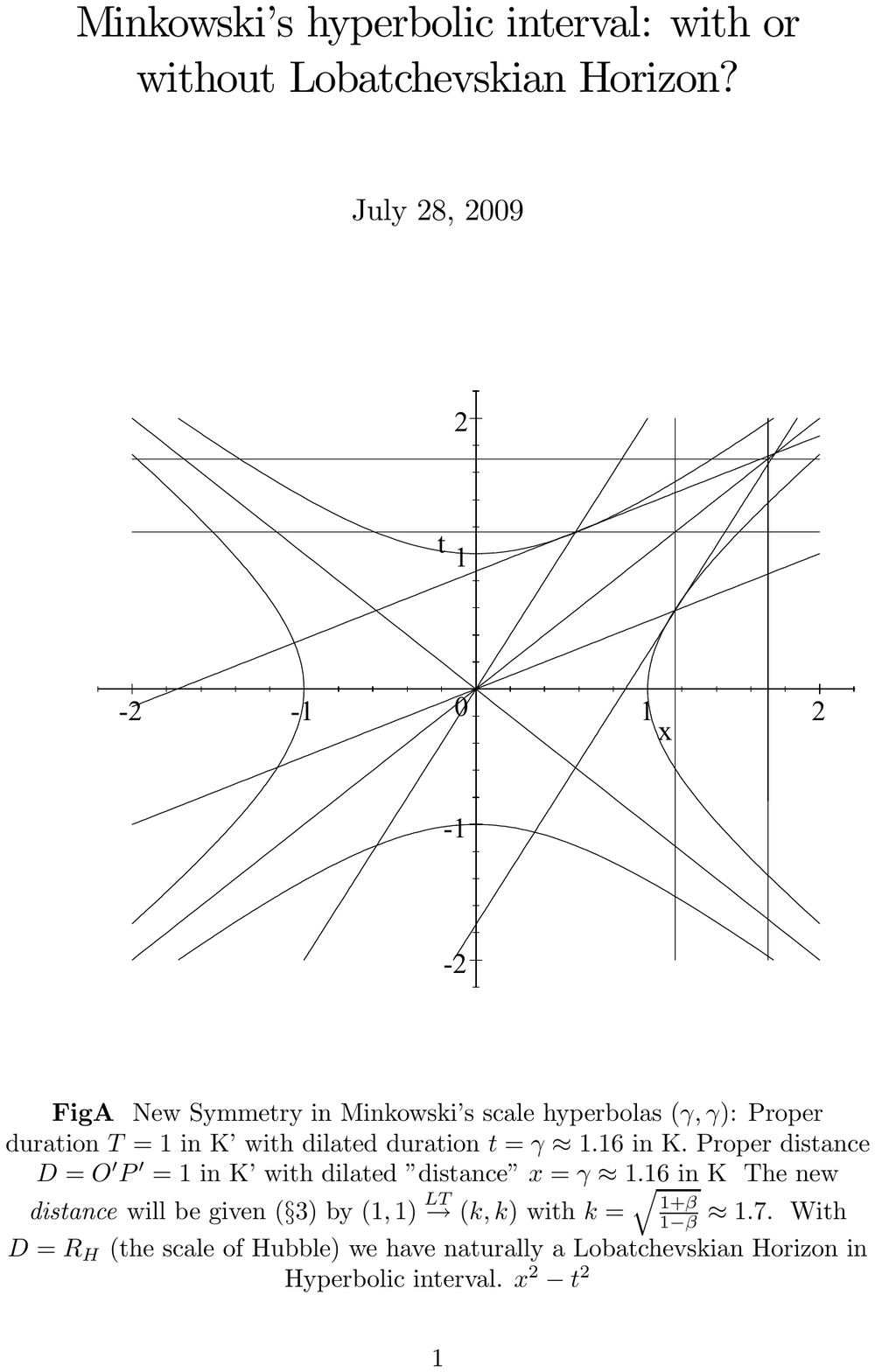} 

\end{document}